# The Kondo Effect in the Unitary Limit

**W.G. van der Wiel[1,*], S. De Franceschi[1], T. Fujisawa[2],**

**J.M. Elzerman[1], S. Tarucha[2,3] and L.P. Kouwenhoven[1]**

[1]*Department of Applied Physics, DIMES, and ERATO Mesoscopic Correlation Project, Delft University of Technology, PO Box 5046, 2600 GA Delft, The Netherlands*

[2]*NTT Basic Research Laboratories, Atsugi-shi, Kanagawa 243-0198, Japan*

[3]*ERATO Mesoscopic Correlation Project, University of Tokyo, Bunkyo-ku, Tokyo 113-0033, Japan*

**We observe a strong Kondo effect in a semiconductor quantum dot when a small magnetic field is applied. The Coulomb blockade for electron tunneling is overcome completely by the Kondo effect and the conductance reaches the unitary-limit value. We compare the experimental Kondo temperature with the theoretical predictions for the spin-1/2 Anderson impurity model. Excellent agreement is found throughout the Kondo regime. Phase coherence is preserved when a Kondo quantum dot is included in one of the arms of an Aharonov-Bohm ring structure and the phase behavior differs from previous results on a non-Kondo dot.**

[*] To whom correspondence should be addressed: email: wilfred@qt.tn.tudelft.nl

The Kondo theory explains the increased resistivity of a metal with magnetic impurities at low temperatures (*1*). Predictions from 1988 indicate that quantum dots could also exhibit the Kondo effect (*2-7*), but now as an increased conductance, *G*, which can reach the unitary limit ($G = 2e^2/h$) at low temperature. Recent experiments have confirmed the presence of the Kondo effect in quantum dots, however not reaching the unitary limit (*8-12*). We demonstrate the unitary limit Kondo effect in a semiconductor quantum dot. Our quantum dot is embedded in one of the arms of an Aharonov-Bohm (AB) ring, which enables us to show that electron transport through the many-body Kondo state is at least partly phase-coherent.

The Kondo effect arises from the coupling between a localized electron spin to a sea of conduction electrons. The strength is characterized by the Kondo temperature, $T_K$ (*13*)

$$T_K = \frac{\sqrt{\Gamma U}}{2} e^{\pi\varepsilon_0(\varepsilon_0+U)/\Gamma U} \tag{1}$$

$U$ is the on-site electron repulsion energy, or charging energy, $\varepsilon_0$ the energy of the single-particle state, and $\Gamma$ reflects its width, due to a finite lifetime from tunneling to the leads (see left inset to Fig. 2A). In quantum dots these parameters can be controlled experimentally, resulting in a "tunable Kondo effect" (*8-12*).

Our device (Fig. 1A) consists of an AB ring defined in a 2-dimensional electron gas (2DEG) (*14*). The conductance of the ring without applying gate voltages is $\sim 10e^2/h$, implying that the current is carried by several modes in each arm. In our experiment, a quantum dot has only been formed in the lower arm. One gate in the upper arm is used to pinch off the upper arm. All measurements are performed in a dilution refrigerator with a base temperature of 15 mK, using a standard lock-in technique with an ac voltage excitation between source and drain contacts of 3 μV.

The linear-response conductance, *G*, through the lower dot versus gate voltage, $V_{gl}$, and magnetic field, *B*, is shown in a color scale plot (Fig. 1B). Here, the left and right parts of the lower arm serve as leads to the dot. In Fig. 1C two $G(V_{gl})$ curves are extracted from Fig. 1B for $B = 0$ and 0.4 T. At $B = 0$, regular Coulomb oscillations are observed with low valley conductance. In some magnetic field ranges, however, the valley conductance increases considerably and can even reach $2e^2/h$, for instance for $B = 0.4$ T. We will discuss Fig. 1B in more detail below, but first focus on the large conductance values observed at $B = 0.4$ T.

Figure 2A shows Coulomb oscillations for different temperatures. At base temperature, the valleys around $V_{gl}$ = -413 mV and -372 mV reach the maximum possible conductance value of $2e^2/h$. In fact, the valleys tend to disappear. When the temperature is increased, two separate Coulomb peaks develop with growing peak spacing. The conductance in the center of the valley has a logarithmic *T*-dependence with a saturation at $2e^2/h$ for low *T*, which is not due to electronic noise (right inset to Fig. 2A). The adjacent Coulomb valleys show an opposite *T*-dependence. This even-odd asymmetry indicates an unpaired spin in a valley with an odd electron number, where we observe the Kondo anomaly, and a spin singlet for an even electron number (*8*,*9*). Fig. 2B shows the differential conductance for different *T* in the middle of the Kondo plateau at $2e^2/h$. The pronounced peak around $V_{SD} = 0$ reflects the Kondo resonance at the Fermi energy. The peak height has the same *T*-dependence as shown in the right inset to Fig. 2A. The width of the peak increases linear with temperature (inset to Fig. 2B).

These measurements are taken after optimizing the two barrier gate voltages, $V_{gl}$ and $V_{gr}$, in order to obtain nearly equal tunnel barriers. However, sweeping $V_{gl}$, as in Fig. 2A, changes the left barrier much more effectively than the right tunnel barrier and hence the barriers cannot be symmetric over the whole $V_{gl}$-range. For a quantitative comparison to theory, we therefore optimize $V_{gr}$ by fixing it at a value chosen such that, upon sweeping $V_{gl}$ , we obtain a flat plateau close to $2e^2/h$ (Fig. 3A) (*15*). The two discernable Coulomb oscillations at higher temperatures have completely merged together at low temperature. This unitary limit was predicted (*3,4*), but not observed before. The unitary limit implies that the transmission probability through the quantum dot is equal to one. This is a remarkable phenomenon since the quantum dot contains two tunnel barriers, each with a transmission probability much less than one. In addition, the on-site Coulomb energy, $U$, tends to block the state with an extra electron on the dot. Despite $U$ being an order of magnitude larger than the characteristic energy scale, $k_BT_K$, the Kondo effect completely determines electron tunneling at low energies (i.e. low $T$ and $V_{SD}$). Note that in the absence of the Kondo effect (e.g. for electron number $N$ = even), the system consists of two separated Fermi seas. In contrast, for $N$ = odd the screening of the local spin creates a single, extended many-body system with a single, well-defined Fermi surface extending throughout the whole system. The quasi-particles at this Fermi surface no longer experience the repulsive barrier potentials, nor the on-site Coulomb repulsion. Also note that since the local spin for $N$ = odd is completely screened and since the dot has zero spin for $N$ = even, the whole system of leads and dot is in a singlet state over a wide gate voltage range (between –430 mV and –350 mV in Fig. 2A), although the nature of the ground state in the even and odd valleys is very different.

For a quantitative analysis we rewrite Eq. 1 as $\ln(T_K) = \pi\varepsilon_o(\varepsilon_o + U)/\Gamma U + constant$, indicating a quadratic dependence for $\ln(T_K)$ on gate voltage, $V_{gl}$ (*16*). Following the work in (*17*), we fit $G$ versus $T$ for different gate voltages (see Fig. 3C) to the empirical function

$$G(T) = G_0\left(\frac{T_K'^2}{T^2 + T_K'^2}\right)^s \tag{2}$$

with $T_K' = T_K/\sqrt{2^{1/s}-1}$ where the fit parameter $s \approx 0.2$ for a spin-1/2 system (*17,18*). Fig. 3B shows the obtained Kondo temperatures, $T_K$ versus $V_{gl}$. The red parabola demonstrates that the obtained values for $T_K$ are in excellent agreement with Eq. 1 (*19*).

The Kondo temperature as derived above, is obtained from the linear response conductance. In earlier works (*8-12*) estimates for $T_K$ were obtained from measurements of $dI/dV_{SD}$ versus $V_{SD}$ ($I$ is the current between source and drain). In that case, the full width at half maximum (FWHM) was set equal to $k_BT_K/e$. However, applying a finite $V_{SD}$ introduces dephasing even at $T$ = 0 (*6,20*). To compare these two methods, we also plot in Fig. 3B (FWHM/$k_B$) measured for different gate voltages at base temperature. Also now we find a parabolic dependence, but the values are larger than $T_K$ obtained from linear-response measurements. The difference may indicate the amount of dephasing due to a non-zero $V_{SD}$.

The normalized conductance, $G/(2e^2/h)$, is expected to be a universal function of the normalized temperature, $T/T_K$, independent of the other energy scales, $U$, $\varepsilon_o$ and $\Gamma$. Over a range of $\Delta\varepsilon_o$ = 225 μeV corresponding to 2.6 Kelvin, which is several times larger than $T_K$, this expected one-parameter scaling is indeed observed in the inset to Fig. 3C.

We now return to the significance of the applied magnetic field. Near $B$ = 0 we observe in Fig. 1B regular Coulomb oscillations. Here, we find that the Kondo effect typically changes the valley conductance by only ~20% (*9*). In Fig. 1B a big change occurs at $B$ ~ 0.1 T, reflecting the onset of a different transport regime, an observation that seems common for half-open quantum dots (*21,22*). The magnetic field scale corresponds to adding a flux quantum to the area of the dot, implying, for instance, that time-reversal symmetry is broken. At $B$ = 0.4 T, where we observe the unitary limit, the Zeeman spin splitting is much smaller than $k_BT_K$ and so it can safely be ignored. This magnetic field scale is also too small for the formation of Landau levels, which can introduce additional spin physics (*22,23*).

Recent calculations (*24*) indicate a spin polarization near $B = 0$ to enhanced values in line with earlier observations (*25,26*). A small magnetic field reduces the spin values to ½ for $N$ = odd and 0 for $N$ = even. Since higher spin states generally lead to a lower Kondo temperature, the Kondo effect can become much stronger by applying a magnetic field (*27*). We believe this to be the origin for the transition in transport regimes near $B = 0.1$ T in Fig. 1B, although a thorough theoretical analysis would be desirable.

So far, the upper arm of the ring was pinched off. To study how electron interference is affected by a Kondo quantum dot in the lower arm, we adjust the upper arm conductance to $\sim 2e^2/h$. In the gray scale plot of Fig. 4A clear AB oscillations around $B = 0.4$ T are visible in the Kondo valley and the neighboring valleys. The period agrees well with a flux quantum, $h/e$, applied through the area enclosed by the ring. The AB oscillations demonstrate that at least part of the tunnel processes through the Kondo quantum dot is phase coherent. Our 2-terminal geometry only allows phase changes by multiples of $\pi$, due to symmetry reasons (*28*). We observe a $\pi$ phase flip at the left Coulomb peak, but no phase change is observed in the Kondo valley or at the right Coulomb peak (Fig. 4B). This behavior is different from the phase evolution described in (*14*). A further study in a 4-terminal geometry should allow us to determine arbitrary phase shifts in the transmission through a Kondo dot, as proposed in (*29*).

30. We thank E. Huizeling, S. Sasaki, D. Goldhaber-Gordon, Y. Meir, M. Eto, L. Glazman, Yu. Nazarov, and R. Schouten for their help. We acknowledge financial support from the Specially Promoted Research, Grant-in-Aid for Scientific Research, from the Ministry of Education, Science and Culture in Japan, from the Dutch Organization for Fundamental Research on Matter (FOM), from the NEDO joint research program (NTDP-98), and from the EU via a TMR network.

**Fig. 1. (A)** Atomic force microscope image of the device. An Aharonov-Bohm (AB) ring is defined in a 2DEG by dry etching of the dark regions (depth is ~75 nm). The 2DEG with electron density, $n_S = 2.6 \cdot 10^{15}$ m$^{-2}$, is situated 100 nm below the surface of an AlGaAs/GaAs heterostructure. In both arms of the ring (lithographic width 0.5 μm, inner perimeter 6.6 μm) a quantum dot can be defined by applying negative voltages to gate electrodes. The gates at the entry and exit of the ring are not used. A quantum dot of size ~200 nm × 200 nm, containing ~100 electrons, is formed in the lower arm using gate voltages, $V_{gl}$ and $V_{gr}$ (the central plunger gate was not working). The average energy spacing between single-particle states is ~100 μeV. The conductance of the upper arm, set by $V_{gu}$, is kept zero except for AB measurements. **(B)** Color scale plot of the conductance, $G$, as function of $V_{gl}$ and $B$ for $V_{gr}$ = -448 mV and $T$ = 15 mK. The upper arm of the AB ring is pinched off by $V_{gu}$ = -1.0 V. Red (blue) corresponds to high (low) conductance. **(C)** Two selected traces $G(V_{gl})$ for $B$ = 0 and $B$ = 0.4 T. The Coulomb oscillations at $B$ = 0 correspond to the oscillating color in (B). For some ranges of $B$, the valley conductance increases considerably, reaching values close to $2e^2/h$, i.e. the unitary limit, e.g. along the yellow dashed line at 0.4 T.

**Fig. 2. (A)** Coulomb oscillations in $G$ versus $V_{gl}$ at $B$ = 0.4 T for different temperatures. $T$ = 15 mK (thick black trace) up to 800 mK (thick red trace). $V_{gr}$ is fixed at –448 mV. The lower-right inset highlights the logarithmic $T$-dependence between ~90 and ~500 mK for $V_{gl}$ = -413 mV. The upper-left inset explains the symbols used in the text with $\Gamma = \Gamma_L + \Gamma_R$. Note that $\varepsilon_o$ is negative and measured from the Fermi level in the leads at equilibrium. **(B)** Differential conductance, $dI/dV_{SD}$, versus dc bias voltage between source and drain contacts, $V_{SD}$, for $T$ = 15 mK (thick black trace) up to 900 mK (thick red trace), also at $V_{gl}$ = -413 mV and $B$ = 0.4 T. The left inset shows that the width of the zero-bias peak, measured from the full-width-at-half-maximum (FWHM) increases linearly with $T$. The red line indicates a slope of 1.7 $k_B/e$, where $k_B$ is the Boltzmann constant. At 15 mK the FWHM = 64 µV and it starts to saturate around 300 mK.

**Fig. 3.** Quantitative analysis in the case of optimized symmetric tunnel barriers. **(A)** $G(V_{gl})$ for $T$ = 15 mK (thick black trace) up to 900 mK (thick red trace) and $V_{gr}$ = –448 mV, $B$ = 0.4 T. **(B)** $T_K$ versus $V_{gl}$ (●) as obtained from many fits as in (C). In addition, we plot the peak width (FWHM/$k_B$) versus $V_{gl}$ (■) as deduced from $dI/dV_{SD}(V_{SD})$ measurements at base temperature (see, e.g., black trace in Fig. 2B). Both data sets are fitted to Eq. 1, resulting in the red and blue parabolas, respectively. **(C)** $G(T)$ at fixed gate voltage as extracted from (A) for $V_{gl}$ = -411 (♦), -414 (○) and –418 (▼) mV (labels are also indicated in (A)). The red curves are fits to Eq. 2. The inset shows that $G$ versus normalized temperature $T/T_K$ scales to a single curve for different gate voltages, $V_{gl}$ = -411 (♦), -412 (□), -413 (×), -414 (○) -415 (+), -416 (▲) mV. The blue curve is a fit to Eq. 2 with fixed $T_K$ = 1 and $G_0 = 2e^2/h$; $s$ = 0.29 is the only fit parameter.

**Fig. 4.** AB oscillations in the conductance through the ring containing the unitary Kondo quantum dot. In this measurement the conductance through the upper arm is set close to $2e^2/h$ in order to have approximately equal transmissions through both arms. **(A)** Gray scale plot of the conductance, $G$, as function of $V_{gl}$ and magnetic field $B$ at 15 mK. Light (dark) corresponds to high (low) conductance. AB oscillations are observed over the whole gate voltage range. The red curve highlights the AB oscillations in the middle of the Kondo plateau ($V_{gl}$ = -412 mV). The period of 0.87 mT corresponds well to a flux quantum, $h/e$, through the area enclosed by the ring. The modulation is 2-3% of the total conductance through the ring. **(B)** Same gray scale plot with the AB conductance maxima indicated by white diamond symbols. A π phase flip is observed when stepping through the left Coulomb peak (e.g. along the red dashed lines). No phase change is observed in the Kondo valley or at the right Coulomb peak.

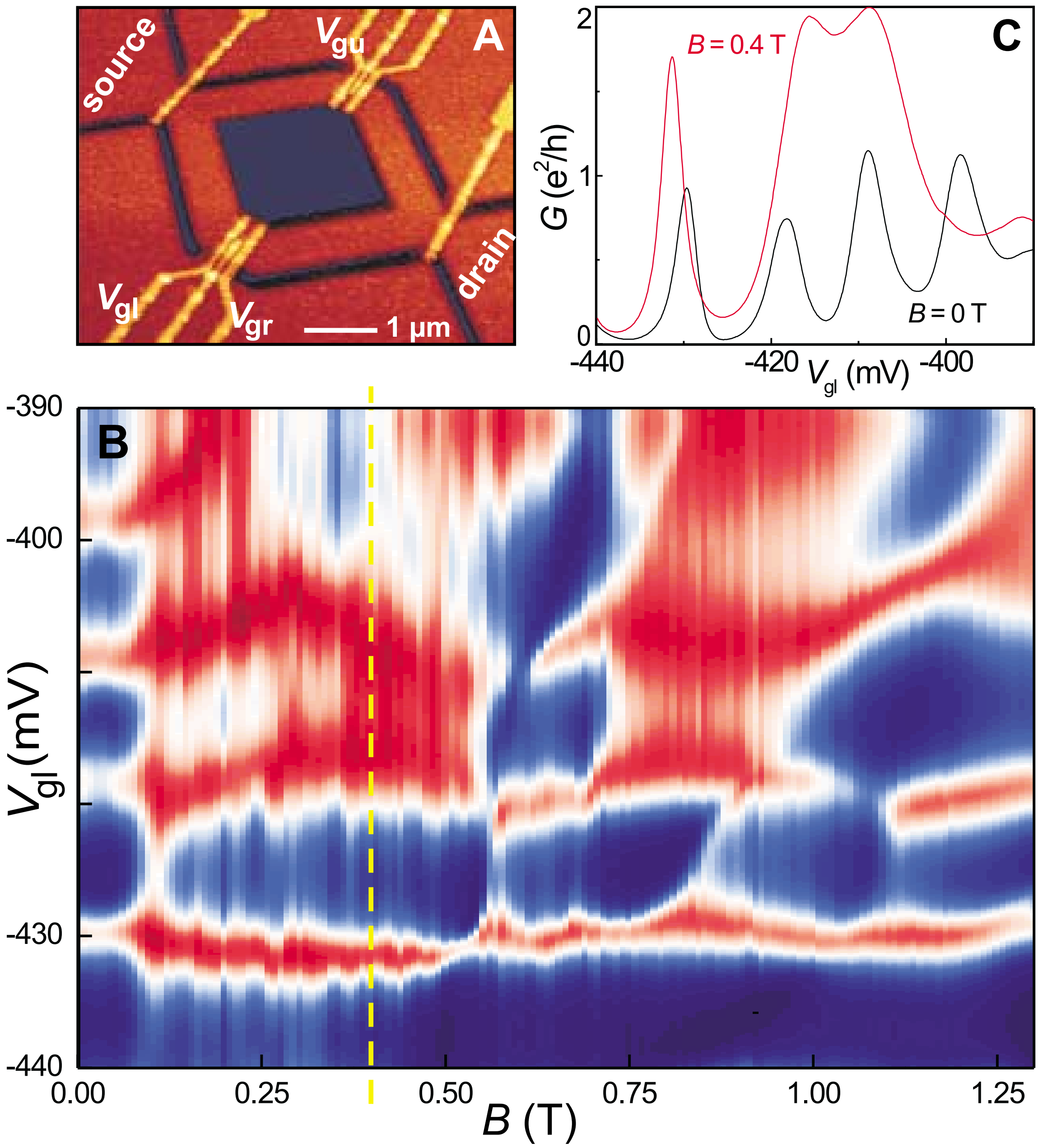

Figure 1
van der Wiel *et al.*

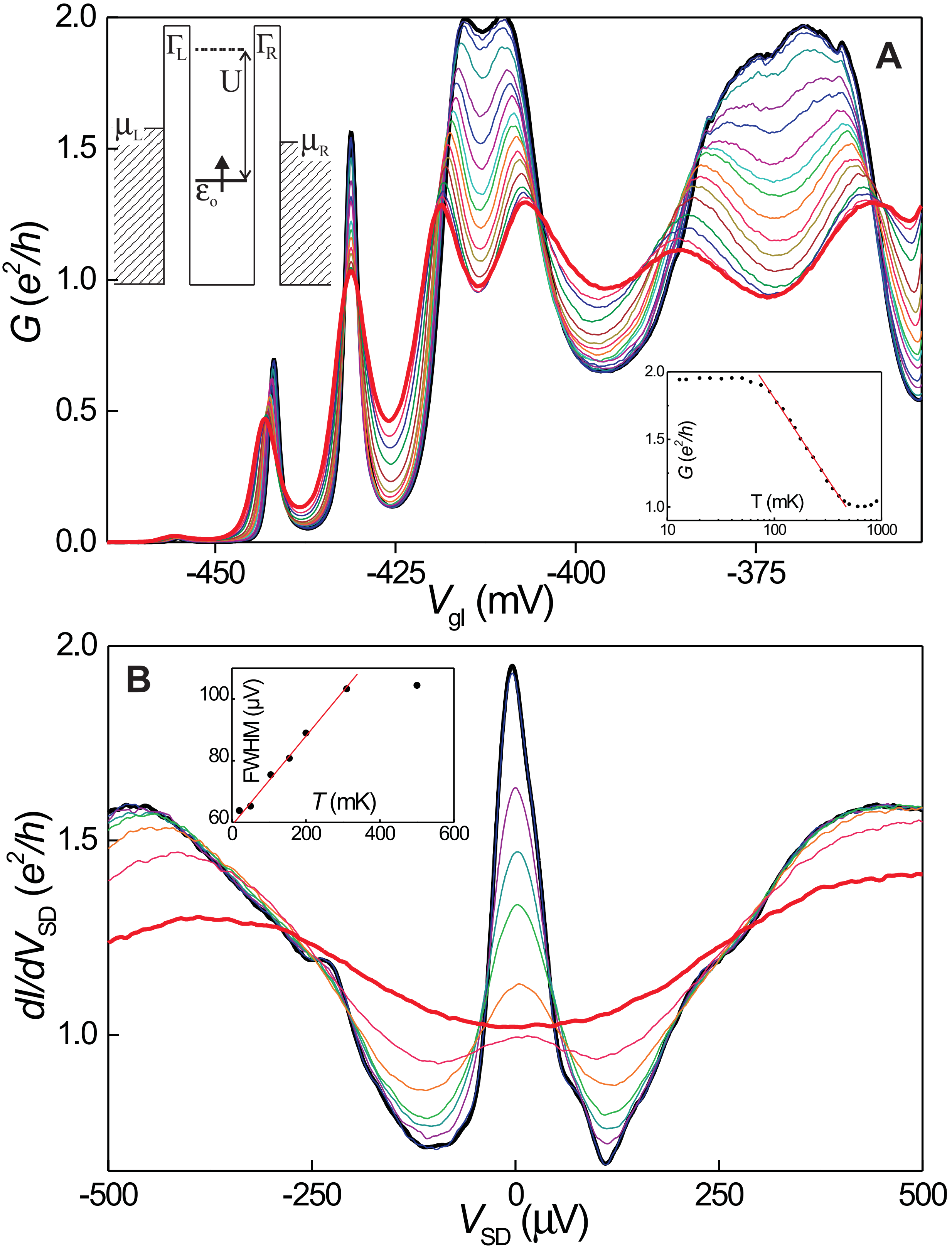

Figure 2
van der Wiel *et al*.

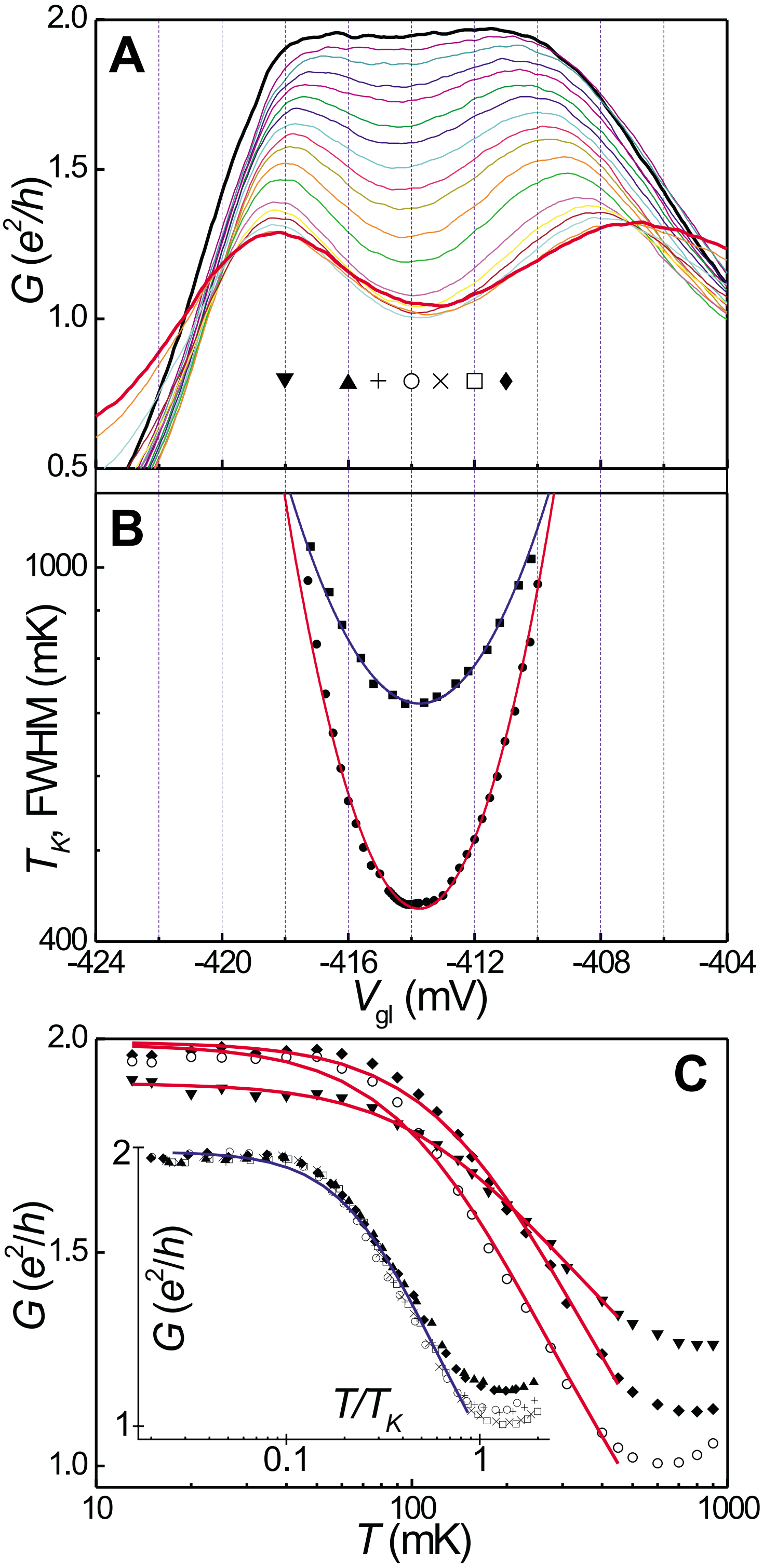

Figure 3 (corrected)
van der Wiel *et al.*

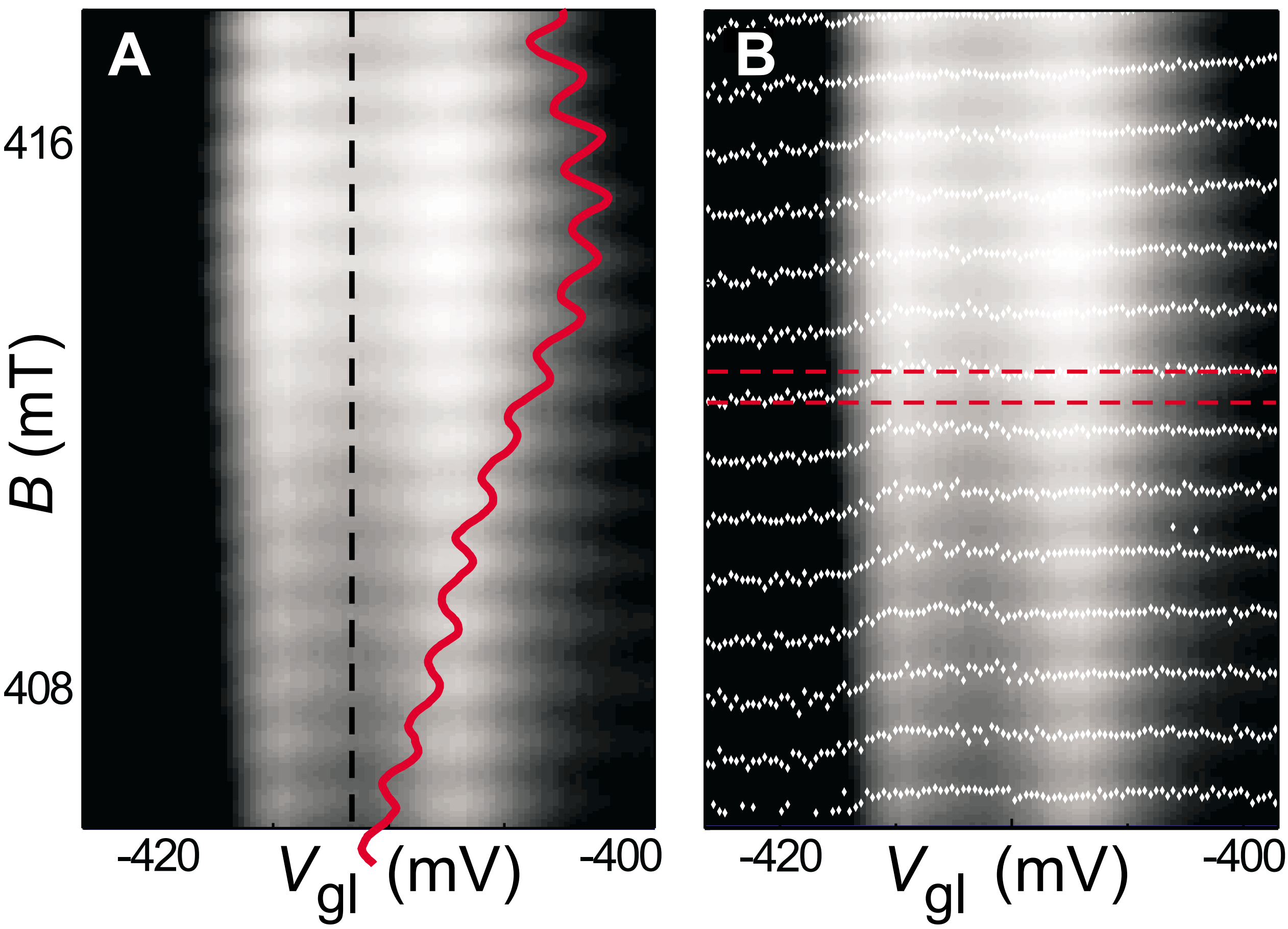

Figure 4
van der Wiel *et al*.